| | |
|---|---|
| Title: | Emittance growth in the DARHT Axis-II Downstream Transport |
| Author(s): | Ekdahl, Carl August Jr.<br>Schulze, Martin E. |
| Intended for: | Report |
| Issued: | 2015-04-14 |

# Emittance growth in the DARHT Axis-II Downstream Transport

Carl Ekdahl and Martin Schulze

*Los Alamos National Laboratory*

*Abstract*

Using a particle-in-cell (PIC) code, we investigated the possibilities for emittance growth through the quadrupole magnets of the system used to transport the high-current electron beam from an induction accelerator to the bremsstrahlung converter target used for flash radiography. We found that even highly mismatched beams exhibited little emittance growth (< 6%), which we attribute to softening of their initial hard edge current distributions. We also used this PIC code to evaluate the accuracy of emittance measurements using a solenoid focal scan following the quadrupole magnets. If the beam is round after the solenoids, the simulations indicate that the measurement is highly accurate, but it is substantially inaccurate for elliptical beams.

I. INTRODUCTION

Flash radiography of hydrodynamic experiments driven by high explosives is a well-known diagnostic technique in use at many laboratories [1, 2]. At Los Alamos, the Dual-Axis Radiography for Hydrodynamic Testing (DARHT) facility provides multiple flash radiographs from different directions of an experiment. Two linear induction accelerators (LIAs) make the bremsstrahlung radiographic source spots for orthogonal views. The 2-kA, 20-MeV Axis-I LIA creates a single 60-ns radiography pulse. The 1.7-kA, 16.5-MeV Axis-II creates multiple radiography pulses by kicking them out of a 1600-ns long pulse from the LIA [3, 4, 5].

Beam emittance is the ultimate limitation on radiographic source spot size. In the absence of beam-target interaction effects, the spot size is directly proportional to the emittance. Since radiographic resolution is limited by the spot size, minimizing emittance enhances resolution of the radiographs. Therefore, investigation and mitigation of factors leading to high emittance beams would be a productive path to improved radiography. In earlier work, we have investigated the potential causes of emittance growth in the DARHT LIAs [6]. In this article, we examine emittance growth in the transport of the beam from the accelerator to the target. This is called the downstream transport (DST).

For a paraxial beam, the normalized emittance is proportional to the volume in phase space, so by Liouville's theorem it is invariant so long as the forces acting on the beam are linear. Although this condition is violated in the fringe fields of focusing magnets, our PIC simulations of the DARHT Axis-II LIA have shown that the normalized emittance of a small, well matched beam is very close to invariant [6]. Likewise, one might expect there to be no growth in the DST. However, distortions of the beam profile, and/or focusing element aberrations can result in nonlinear space-charge and/or focusing forces, and cause to emittance growth, so it is worthwhile pursuing this possibility.

The DST lattice of magnetic focusing elements that transports the beam from the Axis-II LIA to the final focus solenoid incorporates four solenoids and five quadrupoles (Fig.1). The large septum quadrupole is used to divert the un-kicked beam to a dump, and it is followed by four smaller quadrupoles to return the beam to an azimuthally symmetric profile.

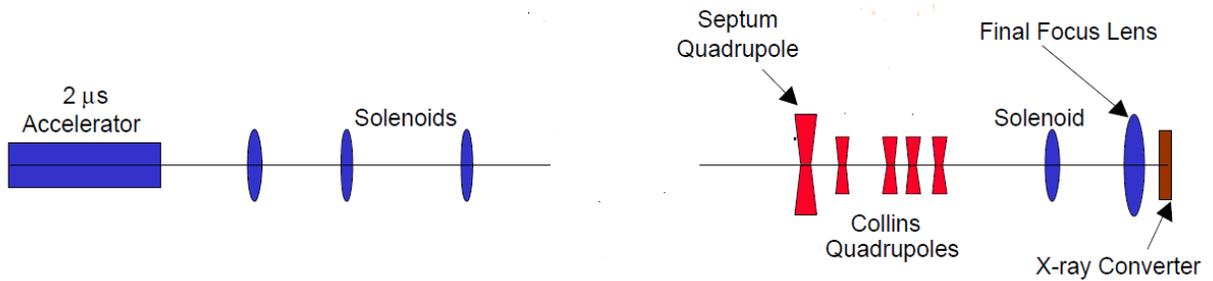

Figure 1: Schematic diagram of the kicked-beam downstream transport showing the four Collins quadrupoles used to return the beam to round following distortion by the septum quadrupole.

Simulations of the DST with the LAMDA envelope code [7] are used to design the tune of these magnets that is used to provide a round focal spot. Figure 2 shows the results of a LAMDA simulation; the curves represent the projections of the hard edge of a uniformly filled ellipse onto the x and y axes, which is the model of the beam density in LAMDA. This figure illustrates how the beam is returned to round by the Collins quads after it has been horizontally focused by passing though the septum quad. Note that the absence of solenoidal fields through the quads implies that the configuration space profile is an upright ellipse until entering the final focus solenoid field.

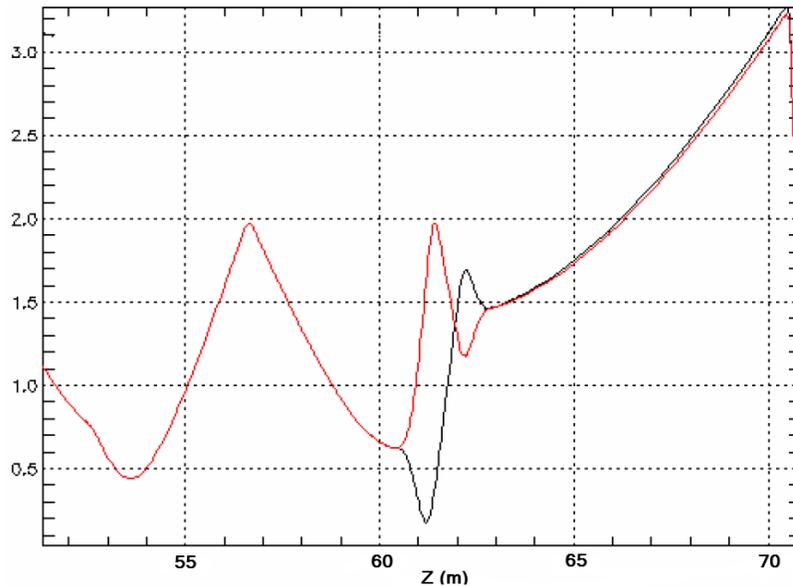

Figure 2: Envelope code simulation of kicked-beam as it is transported through the downstream system. The x and y "envelopes" are the projections of the edge of a uniformly filled ellipse. (red) Y envelope (cm) (black) X envelope (cm). *Adapted from ref.* [5].

The possibility of emittance growth in this downstream transport (DST) system was investigated using a particle-in-cell (PIC) simulation code. The PIC simulations were done in Cartesian coordinates in order to accommodate the 2D focusing properties of the quadrupole magnets. This article is organized as follows. The PIC simulation code and magnetic models used therein are discussed in Section II. The simulation results are discussed in Section III, and some conclusions are presented in Section IV.

## II. PIC SIMULATION CODE

Beam emittance growth in the DST was assessed using a particle-in-cell (PIC) computer code, that is based on the Large Scale Plasma (LSP) code [8]. The LSP-slice algorithm is a simplified PIC model for steady-state beam transport in which the paraxial approximation is assumed [9] A slice of beam particles located at an incident plane of constant z are initialized on a 2D transverse Cartesian (*x,y*) grid. The use of a Cartesian grid admits non-axisymmetric solutions, including beams that are off axis, and transport through quadrupole external fields.

The initial particle distribution of the slice is extracted from a full $x, y, z$ LSP simulation. The initial distribution is a uniform rigid rotor with additional random transverse velocity. The rotation is consistent with zero canonical angular momentum in the given solenoidal magnetic field at the launch position. The random transverse velocity is consistent with the specified emittance.

External fields are input as functions of *z*, and are applied at the instantaneous axial center-of-mass location. External fields that are azimuthally symmetric (fields from solenoids) are input as on-axis values, and the off-axis components are calculated using a power series expansion based on the Maxwell equations [10]. Terms up to fourth order were kept for these simulations.

The on axis magnetic field for the solenoids in the DST was calculated from the magnet models used in our LAMDA envelope code, which are based on magnetic field measurements. The magnetic field for the quadrupoles was input as a map derived from the ideal quadrupole equations,

$$B_x(x, y, z) = g(z)y; B_y(x, y, z) = g(z)x \qquad (1.1)$$

where *g(z)* is the field gradient. The quadrupole gradients used to generate these maps were taken from the quadrupole models used in LAMDA, which are based on measurements for each of the magnets. That is, the LAMDA input *.b3d files of gradients on axis were converted to *.m3d maps for LSP using a purpose-built IDL program. Thus, the magnetic fields used in the PIC simulations are derived from the same physical measurements as those used in LAMDA envelope code calculations, which helps comparison of results between the two codes.

It is worth noting that both the LAMDA and XTR [11] envelope codes and the LSP PIC code specify beam energy in terms of "beam voltage" or "wall potential" with units of MV. This is the sum of accelerating potentials such as the diode voltage and cell voltages. The actual kinetic energy of beam electrons is this wall potential less the beam space-charge depression. This is exact in the PIC code for a given beam pipe size, but only approximate in the envelope codes.

## III. SIMULATION RESULTS

A. *Comparison with envelope codes*

Previous LSP-Slice PIC simulations of round beams in the solenoidal fields of the DARHT LIAs have agreed with simulations by the envelope code XTR [4] [6] [12]. The version of XTR used for these comparisons has exactly the same physics models and approximations for round beams as the LAMDA envelope code, which can also simulate elliptical beams transported by quadrupole magnets. Moreover, LAMDA has been used to develop the tunes for the DST that are used in practice. Therefore, it is reasonable to compare our PIC simulations of the DST with LAMDA as a reality check.

On the other hand, there are some differences between the LAMDA physics models and the natural physics of the PIC code. The most notable of these are the following:

- LAMDA (and XTR) accepts pipe sizes that vary with position; LSP-Slice does not. This has a slight effect on the space-charge depression of the beam kinetic energy. It was not an issue in LIA simulations, because they were done in regions where the pipe size was constant.
- The envelope of an elliptical uniform beam is not an equipotential, and LAMDA uses a simple approximation for the space-charge depression; that it is the same as for a round beam with an envelope size equal to the average of the semi-minor and semi-major dimensions of the ellipse.
- LAMDA does not correct the external fields for beam diamagnetism in the elliptical beam models. Therefore, to reduce differences due to transport in solenoidal fields, we launch the PIC beam further down the DST than the initial position typically used for LAMDA in order to minimize time spent in solenoidal fields.
- Magnetic focusing is different in the two codes. In LAMDA the focusing is derived from the axial magnet field and magnetic gradients on axis, while the PIC code calculates the focusing of electrons from field maps.
- The normalized emittance in LAMDA and XTR is assumed to be invariant (a number provided by the user), because the external forces are linear, and space charge forces from a uniform distribution are also linear for a round beam. However, this is not true of an elliptical distribution, so it is inconsistent. On the other hand, the PIC code calculates it from actual distributions in phase space at each step of the propagation, using non-linear external and space charge forces. Thus it need not be invariant in the PIC code.

The results of a recent LAMDA simulation of the DST are shown in Fig. 3. The initial beam parameters for this simulation are given in Table I. The magnet settings used are given in Table II.

Table I. Initial beam parameters for LAMDA simulation

| Parameter | Symbol | Units | Value | Comment |
|---|---|---|---|---|
| Initial Position | $z_0$ | cm | 5050.0 | LIA Exit |
| Wall Potential | $\phi$ | MeV | 17 | KE=phi-space charge depression |
| Current | $I_b$ | kA | 2 | |
| Emittance (normalized) | $\varepsilon_n$ | cm-radian | 0.086 | |
| Envelope Radius | $r_0$ | cm | 1.11 | |
| Convergence | $r_0'$ | mr | 3.85 | convergence => positive |
| External Field | $B_z$ | G | 0.0 | |

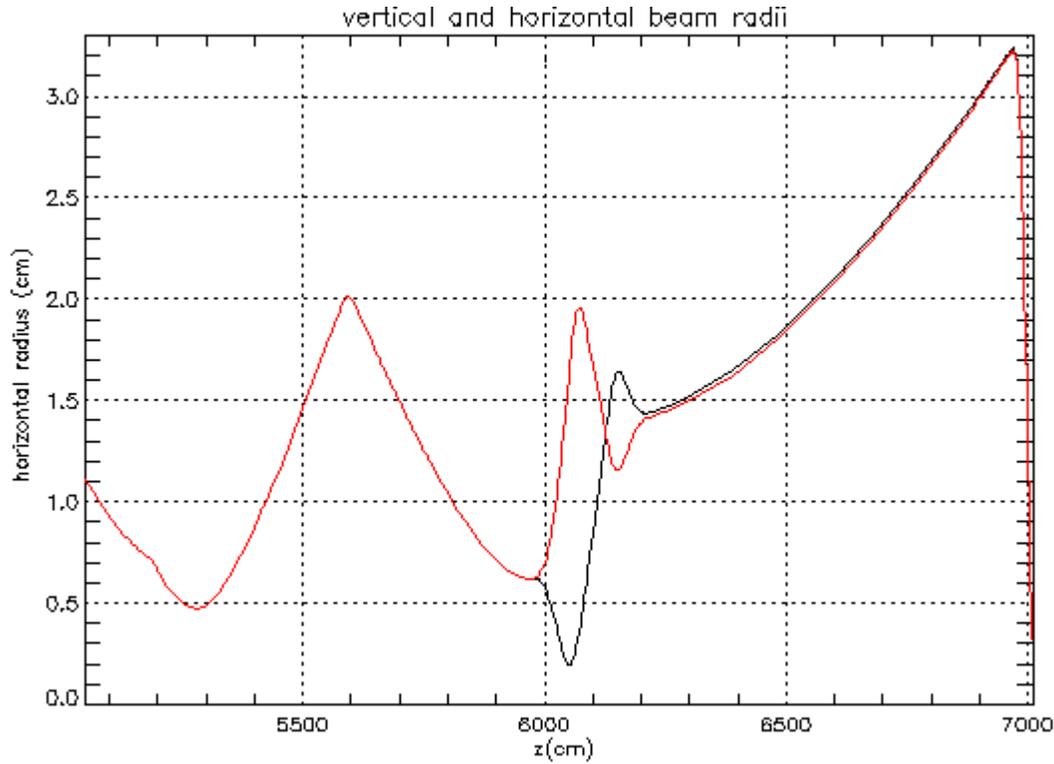

Figure 3: Elliptical-beam envelope calculated in a recent LAMDA simulation corresponding to Fig. 2. The x and y "envelopes" are the projections of the edge of a uniformly filled ellipse. (red) Y edge (cm) (black) X edge (cm).

Table II. Magnets in the DARHT Axis-II downstream transport (DST) system

| Magnet | Type | Descriptor | LSP File Name | Location cm | Current (LAMDA) A | Current (LSP-PIC) A |
|---|---|---|---|---|---|---|
| S1 | Solenoid | Post-LIA | s1mea.m3d | 5188.902 | 64.0 | N/A |
| S2 | Solenoid | Cruncher | s2mea.m3d | 5414.234 | 0.0 | N/A |
| S3 | Solenoid | Pre-Kicker | s3mea.m3d | 5592.474 | 75.0 | 75.0 |
| SQ | Quadrupole | Septum | qsmea.m3d | 6007.949 | -120.0 | -120.0 |
| CQH | Quadrupole | Collins | cqhmea.m3d | 6067.755 | 60.3 | 60.1 |
| CQV | Quadrupole | Collins | cqvmea.m3d | 6145.532 | -45.2 | -44.7 |
| CQW | Quadrupole | Collins | cqwmea.m3d | 6196.344 | 12.0 | 12.7 |
| CQX | Quadrupole | Collins | cqxmea.m3d | 6245.797 | 0.0 | 0.0 |
| S4 | Solenoid | Scan Solenoid | s4mea.m3d | 6604.221 | 0.0 | 0.0 |
| SFF | Solenoid | Final Focus | ffmea.m3d | 6983.286 | 486.0 | 486.0 |

The LSP-Slice PIC simulations included only the last 7 DST magnets beginning with solenoid S3. These are listed in Table II. The initial position was at the envelope maximum near the center of the S3 solenoid. Initial beam parameters for the PIC simulations are listed in Table III. Also, for the S4 solenoid turned off, there is a 684-cm field free drift region between 6219 cm and 6903 cm. The strong field of the final focus solenoid has a profound effect on the large beam, through Larmor rotation and non-linear external forces. Therefore, this exploration of emittance growth and other effects is confined to the space between the initial position in S3 and the beginning of the final focus field in order to minimize time in axial magnetic fields.

Figure 4 shows the results from a PIC simulation in which the Collins quadrupole settings were slightly varied from the LAMDA values to obtain the nearly round beam in the post-quad drift region (see Table II). We speculate that this was needed to account for the previously mentioned differences between physics models in the codes. As seen in Fig. 4, the emittance for this baseline tune is almost constant throughout the DST, with less than 5% growth. This is likely due to the slight softening of the initial hard-edge distribution. This softening is clearly evident in movies of the beam distribution.

Table III. Initial beam parameters for LSP-Slice PIC code simulations.

| Parameter | Symbol | Units | Value | Comment |
|---|---|---|---|---|
| Initial Position | $z_0$ | cm | 5593.75 | |
| Wall Potential | $\phi$ | MeV | 17 | KE=phi-space charge depression |
| Current | $I_b$ | kA | 2 | |
| Emittance (normalized) | $\varepsilon_n$ | cm-radian | 0.086 | |
| Radius | $r_0$ | cm | 2.029 | |
| Convergence | $r_0'$ | mr | 0.0 | |
| External Field | $B_z$ | G | 1875.0 | |

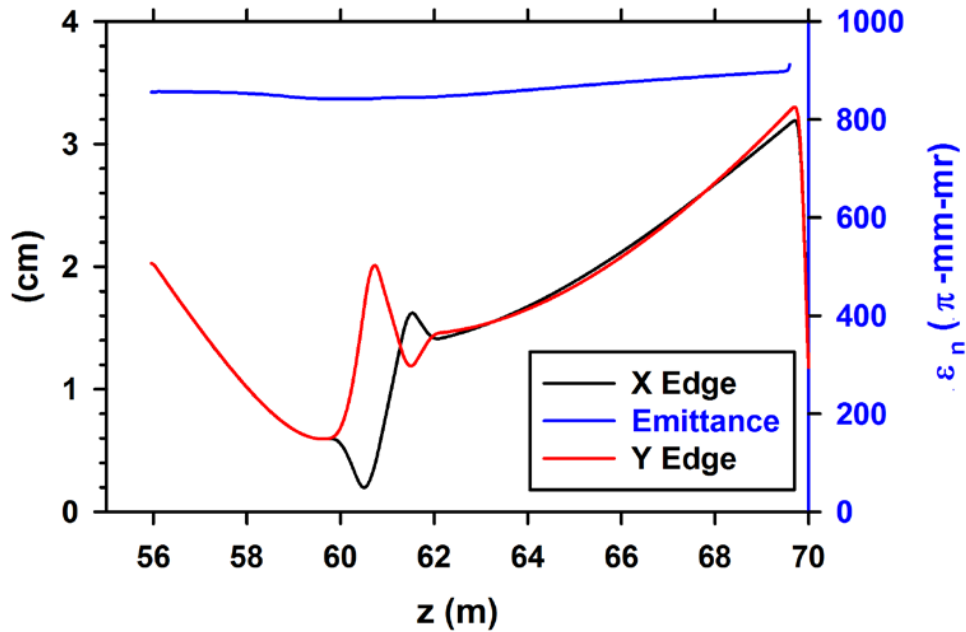

Figure 4: The x and y edges (envelope projection) are calculated from rms values of projections of the distribution. (red) Y edge (cm) (black) X edge (cm). Also shown is the normalized emittance in (blue).

B. *Emittance growth*

It might be expected that the emittance would grow in the DST if the initial conditions were other than those given above. This was examined for round beams in the LIAs by varying the initial energy to cause a mismatch. This is a useful means for investigating mismatches of beam initial conditions to tunes of the DST, in part because the energy is slightly different for the four pulses. Thus, this approach can explore differences in beam quality between pulses with different energies; as in actual operations in which the pulse energies are varied by varying the timing of the last two or three cells of the LIA. Completely turning off a single cell depresses the beam energy by ~230 keV (200-kV accelerating potential plus 30-kV beam loading). Of course, dropping the initial energy changes the beam size at the slice launch position, and this change was determined with XTR. We also investigated the beam quality degradation by a complete failure of a single block of six cells.

Large elliptical radiographic source spots are undesired. Therefore, it is useful to evaluate the effects of beam mismatch using metrics based on beam ellipticity after traversing the DST, and beam emittance, which determines spot size in the absence of beam-target interactions.

Thus, a measure of the quality of the tune is the ellipticity of the beam in the drift region between the last quadrupole and the final focus solenoid. This can be quantified by the ellipse flattening parameter, defined as $f = 1 - a/b$, where $a$ is the semi-minor axis, and $b$ is the semi-major axis, so that $f \leq 1$. Smaller is better, and a round beam has $f = 0$.

Another measure of quality is the emittance growth between the initial position and 6903 cm, where the beam enters the field of the final focus solenoid. These are given in Table IV for the baseline tune (Fig. 4), variations in energy corresponding to turning off one or two cells in the LIA, and catastrophic failure of several cells.

Table IV. Results of PIC simulations of mismatched beam

| Wall Potential $\phi$ | Initial Radius $r_0$ | Average Flatness $\langle f \rangle$ | Maximum Flatness $f_{max}$ | Final Flatness $f_{final}$ | Final emittance $\varepsilon_{n\,final}$ | emittance growth $\delta\varepsilon_n$ | Comment |
|---|---|---|---|---|---|---|---|
| MV | cm | | | | mm-mr | % | |
| 17.00 | 2.03 | 0.0131 | 0.0294 | 0.0215 | 895 | 4.1 | baseline |
| 16.77 | 2.05 | 0.0132 | 0.0578 | 0.0225 | 896 | 4.2 | one cell off |
| 16.54 | 2.07 | 0.0207 | 0.0866 | 0.0044 | 896 | 4.2 | two cells off |
| 15.62 | 2.18 | 0.2015 | 0.2825 | 0.2655 | 907 | 5.5 | 6-cell failure |

From the results presented in Table IV it is clear that turning one or two cells off has little effect on beam quality at the final focus (final flatness and emittance). In fact, with two cells off the beam is closer to round entering the final focus than the baseline tune at full energy. On the other hand, the failure of a cell block would have a disastrous effect on the radiographic spot. In this case the flatness at the final focus would be more than an order of magnitude greater than at full energy.

C. *Emittance measurements*

Emittance measurements are essential for improving beam quality, thereby improving the resolution of DARHT radiographs. An appropriate beam optics code can then be used to find the beam initial conditions at an upstream point by maximum likelihood fitting to the data [13] [6]. In our most recent measurements we used a solenoid 3.8 m upstream of the final focus to change the size of 50-ns beam pulse produced by the kicker. We imaged the optical transition radiation (OTR) from a 51-micron thick Ti target with a 10-ns gated camera. We used the XTR envelope code to fit our data to find the beam envelope size, divergence, and emittance at a position upstream of the focusing solenoid. It can be shown that the sensitivity of the image spot size to emittance is maximized by increasing the drift distance from the initial position to the focusing magnet. Therefore, for the analysis of many of our measurements we chose this position to be 6246 cm (3.58 m upstream from the solenoid) [6], which is well within the field-free drift region, and certainly not influenced by the focusing magnet.

We used LSP-Slice to investigate the accuracy of this measurement technique. In a series of simulations we varied the S4 focusing magnet and obtained the rms value of the PIC distribution at the imaging target location. From these, we calculated the value of the edge of the projection of an equivalent uniform density ellipse (edge=2 x rms). The XTR envelope code was then used to fit the curve of edge values vs focusing magnet current, and the parameters from the best fit compared with the actual PIC values to get6 an estimate of the uncertainty in the final values.

Of course, this is a highly idealized estimate of uncertainty, and includes none of the complications of real experimental measurements of real beam distributions [6] [13]. Nevertheless, it does provide some insight into the best that one can expect from this technique. These simulations were done for both the baseline, matched beam ($\phi = 17.0\text{MV}, r_0 = 2.03\text{ cm}$) and the mismatched beam ($\phi = 15.62\text{MV}, r_0 = 2.18\text{ cm}$) which was highly elliptical downstream of the focusing magnet. Figure 5 shows the rms sizes that were extracted from the PIC results for the matched and mismatched beams. The results of XTR fitting to the PIC results, and the errors in the results of the fits are listed in Table V.

Figure 6 shows the equivalent *x* and *y* beam envelope edges calculated from the rms sizes for the matched beam ($x_{edge} = 2x_{rms}, y_{edge} = 2y_{rms}$), and the XTR fits to those edges. Since the matched beam is close to round for the entire distance from initial position to imaging target, the XTR fit is good, and the error in this simulated emittance measurement is small, $< 2\%$.

On the other hand, the mismatched beam is highly elliptical, and as shown in Fig. 7 the XTR fits are poor, with large errors in the emittance, $> 30\%$. In practice, we have tried to account for beam ellipticity by using a size that is the average of the projections into many angles. Therefore, we included in this analysis an attempt to fit the average of the edges shown in Fig. 7. As seen in Fig. 8, the XTR fit was excellent, but the error in emittance was still very large, >30%. That is, XTR provided an excellent fit to incorrect initial conditions.

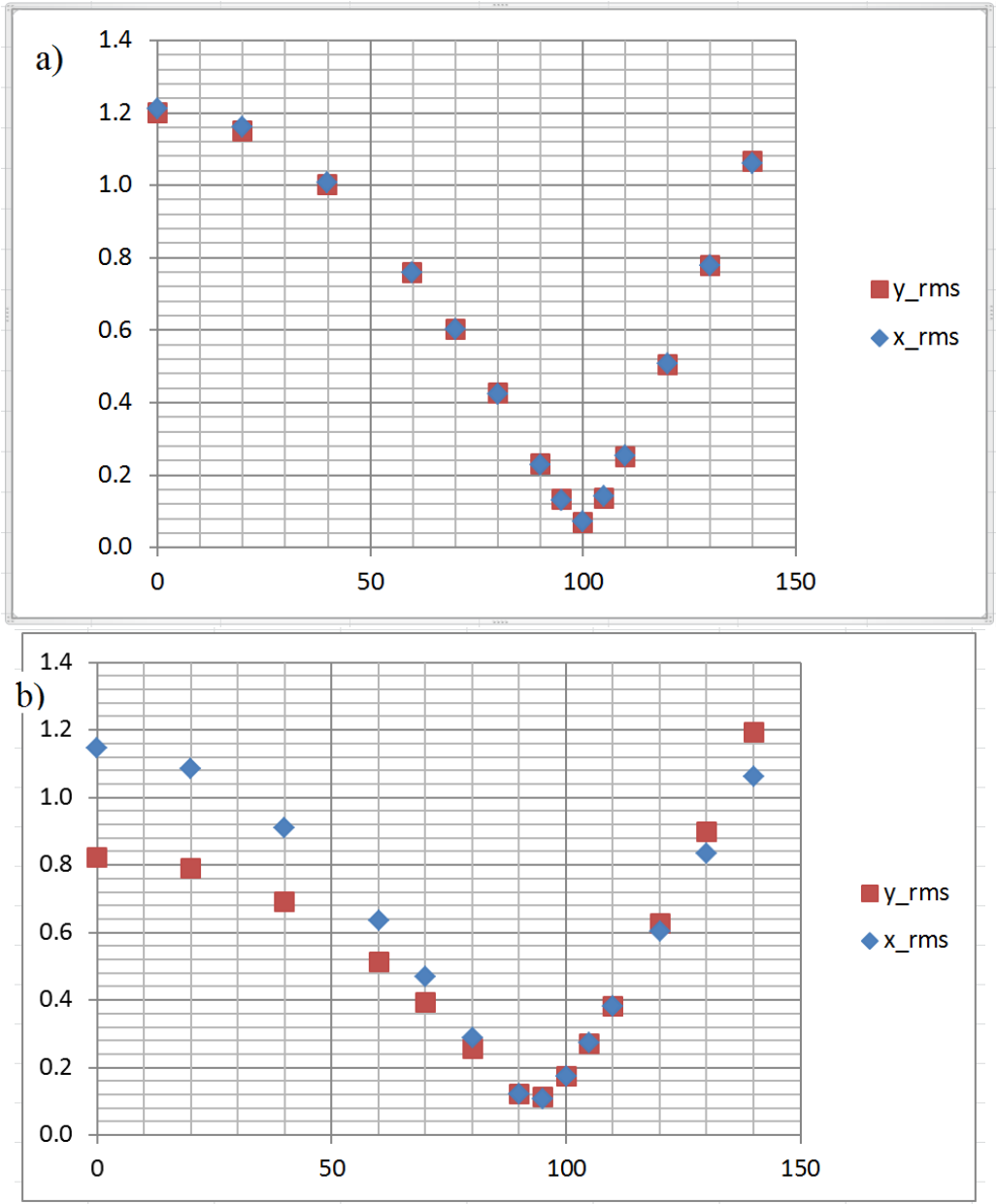

Figure 5: PIC code results for beam size at location of imaging target. a) Baseline beam parameters; $\phi = 17.0\,\text{MV}$, $r_0 = 2.03\,\text{cm}$ and Table III. b) Mismatched beam parameters; $\phi = 15.62\,\text{MV}$, $r_0 = 2.18\,\text{cm}$ and Table III.

Table V. Initial conditions calculated by XTR envelope code by fitting PIC results of varying S4 solenoid.

| | | | Value | Error (%) |
|---|---|---|---|---|
| Locations | | | | |
| | XTR Initial Position | $z_0$ (cm) | 6245.797 | |
| | Focusing solenoid | $z_{S4}$ (cm) | 6604.221 | |
| | Imaging Target | $z_C$ (cm) | 6712.915 | |
| Initial Conditions | | | | |
| | Envelope Edge, X | $x_e$ (cm) | 1.45 | |
| | Envelope Edge, Y | $y_e$ (cm) | 1.48 | |
| | Emittance | $\varepsilon_n$ (π-mm-mr) | 849 | |
| XTR Results, Baseline | | | | |
| | Envelope Edge, X | $x_e$ (cm) | 1.63 | 12.8 |
| | Envelope Edge, Y | $y_e$ (cm) | 1.61 | 8.3 |
| | Emittance, X | $\varepsilon_n$ (π-mm-mr) | 838 | 1.3 |
| | Emittance, Y | $\varepsilon_n$ (π-mm-mr) | 834 | 1.8 |
| XTR Results, Mismatched | | | | |
| | Envelope Edge, X | $x_e$ (cm) | 2.49 | 97.7 |
| | Envelope Edge, Y | $y_e$ (cm) | 2.44 | 69.5 |
| | Emittance, X | $\varepsilon_n$ (π-mm-mr) | 1204 | 49.6 |
| | Emittance, Y | $\varepsilon_n$ (π-mm-mr) | 1131 | 31.0 |
| | | | | |
| | Mean Envelope Edge | $\bar{r}_e$ (cm) | 2.45 | 81.3 |
| | Emittance | $\varepsilon_n$ (π-mm-mr) | 1154 | 33.7 |

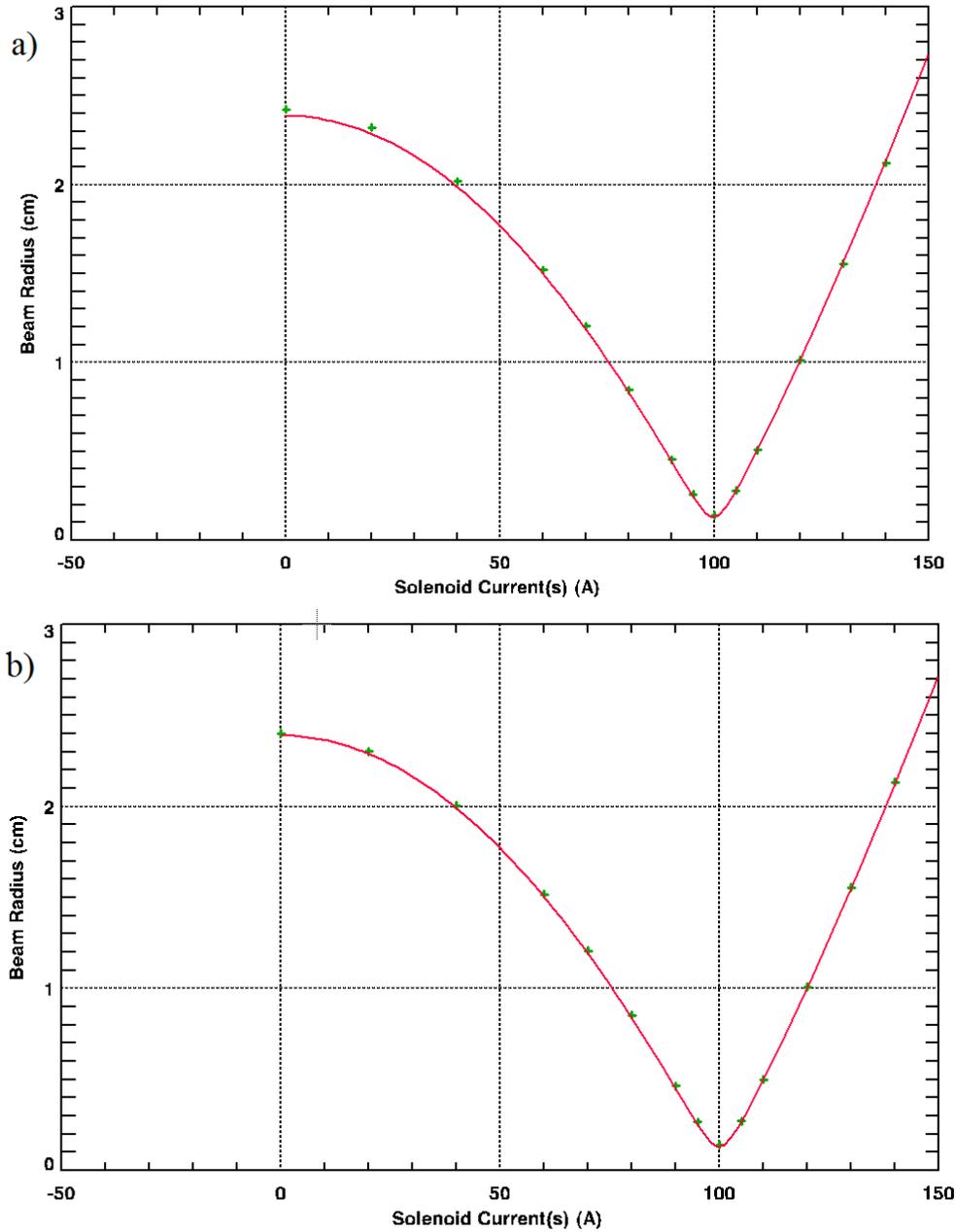

Figure 6: XTR envelope code fits to edge envelopes derived from PIC results for baseline beam parameters shown in Fig. 2a. a) X edge. Best fit initial values at 3.58 m upstream of focusing magnet; $r_0 = 1.63$ cm, $\varepsilon_n = 838$ $\pi$-mm-mr   b) Y edge. Best fit initial values at 3.58 m upstream of focusing magnet; $r_0 = 1.60$ cm, $\varepsilon_n = 834$ $\pi$-mm-mr.

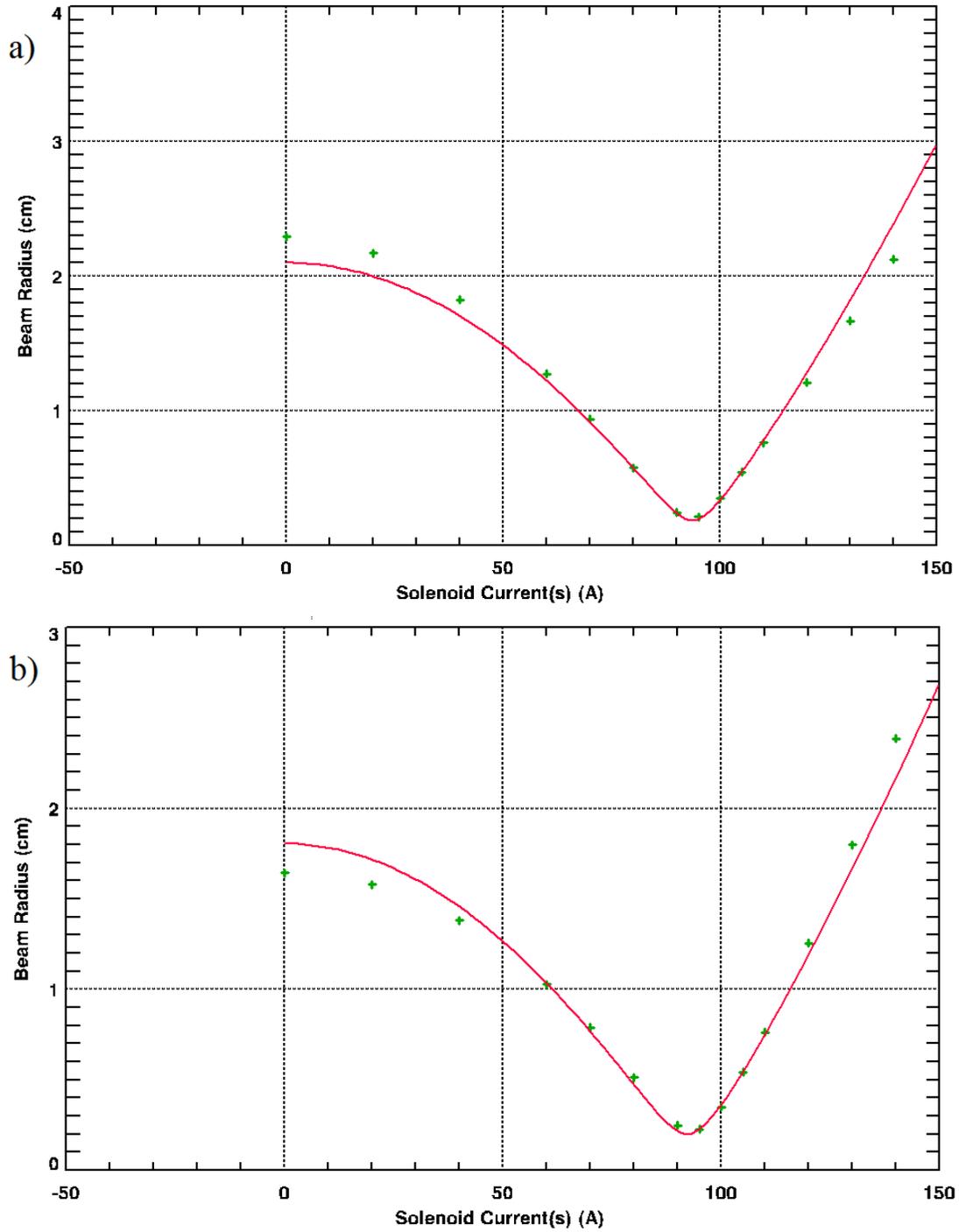

Figure 7: XTR envelope code fits to edge envelopes derived from PIC results for mismatched beam parameters shown in Fig. 2b. a) X edge. Best fit initial values at 3.58 m upstream of focusing magnet; $r_0 = 2.49$ cm, $\varepsilon_n = 1204$ $\pi$-mm-mr    b) Y edge. Best fit initial values at 3.58 m upstream of focusing magnet; $r_0 = 2.44$ cm, $\varepsilon_n = 1131$ $\pi$-mm-mr.

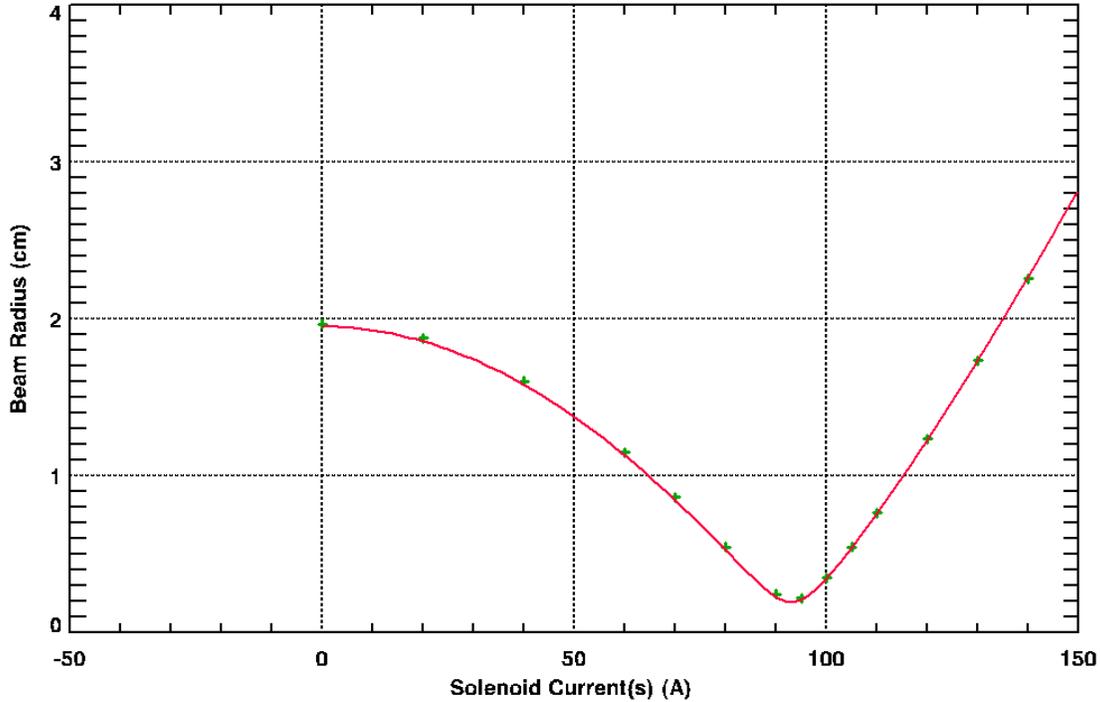

Figure 8: XTR envelope code fits to edge envelopes derived from PIC results for mismatched beam parameters shown in Fig. 2. The radius was approximated by averaging the x and y projection edges.

IV. CONCLUSIONS

Emittance growth in the DARHT Axis-II downstream transport system is insignificant when it is well tuned to produce a round beam after the quadrupole magnets. The ~4% growth observed is likely due to the softening of the initial hard-edge distribution observed in movies of the evolution of the beam distribution as it is transported through the DST.

Emittance and ellipticity are scarcely affected by turning one or two cells off. However, failure of an entire cell block causes a mismatch of the beam to the DST that would have a deleterious effect on beam quality and the radiographic spot.

PIC simulations of emittance measurements by the focal scan technique show excellent agreement between results and simulated beam parameters for beams well matched to the tune of the DST magnets. However, significant errors were found in the results for mismatched beams having highly elliptical distributions downstream of the quads. Moreover, using average sizes also produced incorrect results, especially misleading because of the visibly good fit by XTR to the averaged data This emphasizes the necessity for tuning the DST to produce a round beam at all settings of the focusing magnet before attempting to measure the emittance with this technique.

Acknowlegements

The authors once again acknowledge that this investigation would not have been feasible without the fast-running LSP-Slice algorithm developed by Carsten Thoma and Tom Hughes [9]. We also thank all of our colleagues at the Los Alamos DARHT for invigorating and enlightening discussions.

This research was supported by the US Department of Energy and the National Nuclear Security Administration under contract number DE-AC52-06NA253960.